\documentclass[a4paper, 11pt]{article}
\usepackage{comment} 
\usepackage{lipsum} 
\usepackage{fullpage} 
\renewcommand{\thesubsection}{\alph{subsection}}
\usepackage{booktabs}
\usepackage{setspace}
\usepackage{everypage}
\usepackage{environ}
\usepackage[hidelinks]{hyperref}
\usepackage{booktabs}
\usepackage[english]{babel}
\usepackage[utf8x]{inputenc}
\usepackage[margin=1in]{geometry}
\usepackage{graphicx}
\usepackage{subcaption}
\graphicspath{{desktop/}}
\usepackage[colorinlistoftodos]{todonotes}
\usepackage{amsmath,amssymb,amsthm}
\usepackage{bm}
\usepackage{mathrsfs}
\usepackage[colorinlistoftodos]{todonotes}
\usepackage[export]{adjustbox}
\usepackage{setspace}
\usepackage{natbib}
\setcitestyle{aysep={,}} 
\usepackage{pdflscape}
\usepackage{booktabs,caption}
\usepackage[flushleft]{threeparttable}
\usepackage{afterpage}
\renewcommand\thesection{\arabic{section}}
\renewcommand\thesubsection{\thesection.\arabic{subsection}}
\usepackage[toc,page]{appendix}
\usetikzlibrary{arrows.meta}
\newtheorem{condition}{Condition}
\newcounter{abspage}

\makeatletter
\newcommand{\newSFPage}[1]
  {\global\expandafter\let\csname SFPage@#1\endcsname\null}

\NewEnviron{SidewaysFigure}{\begin{figure}[p]
\protected@write\@auxout{\let\theabspage=\relax}
  {\string\newSFPage{\theabspage}}%
\ifdim\textwidth=\textheight
  \rotatebox{90}{\parbox[c][\textwidth][c]{\linewidth}{\BODY}}%
\else
  \rotatebox{90}{\parbox[c][\textwidth][c]{\textheight}{\BODY}}%
\fi
\end{figure}}

\AddEverypageHook{
  \ifdim\textwidth=\textheight
    \stepcounter{abspage}
  \else
    \@ifundefined{SFPage@\theabspage}{}{\global\pdfpageattr{/Rotate 0}}%
    \stepcounter{abspage}%
    \@ifundefined{SFPage@\theabspage}{}{\global\pdfpageattr{/Rotate 90}}%
  \fi}
\makeatother

\newtheoremstyle{indented}
{3pt}
{3pt}
{\itshape}
{2.5em}
{\bfseries}
{:}
{.5em}
{\thmname{#1}\thmnumber{ #2}\thmnote{ (#3)}}

\theoremstyle{indented}
\newtheorem{thm}{Theorem}

\theoremstyle{indented}

\title{
Interpreting Instrumental Variable Estimands with Unobserved Treatment Heterogeneity: The Effects of College Education%
\footnote{I thank Haiqing Zhao, who was a coauthor on a related paper (Estimating Returns to Unobserved Education, 2022), for giving extensive feedback. I also thank 
Xiaoxiao Li, Anna Mikusheva, Jeff Smith, Chris Taber, and Matt Wiswall, 
as well as seminar participants at the Midwest Economics Association Annual Meeting, the Society of Labor Economists Annual Meeting, the Southern Economic Association Annual Meeting, and the University of Wisconsin-Madison for helpful comments. This research was conducted with restricted access to Bureau of Labor Statistics (BLS) data. The views expressed here do not necessarily reflect the views of the BLS.
}
}

\author{Clint Harris\footnote{University of Wisconsin-Madison, 330 N Orchard Street,
Madison, WI 53715, USA; email: clint.harris@wisc.edu}
}

\date{\today}

\begin{document}
\maketitle

\begin{abstract}
Many treatment variables used in empirical applications nest multiple unobserved versions of a treatment. I show that instrumental variable (IV) estimands for the effect of a composite treatment are IV-specific weighted averages of effects of unobserved component treatments. Differences between IVs in unobserved component compliance produce differences in IV estimands even without treatment effect heterogeneity. 
I describe a monotonicity condition under which IV estimands are positively-weighted averages of unobserved component treatment effects. 
Next, I develop a method that allows instruments that violate this condition to contribute to estimation of treatment effects by allowing them to place nonconvex, outcome-invariant weights on unobserved component treatments across multiple outcomes. 
Finally, I apply the method to estimate returns to college, finding wage returns that range from 7\% to 30\% over the life cycle. 
My findings emphasize the importance of leveraging instrumental variables that do not shift individuals between versions of treatment, as well as the importance of policies that encourage students to attend ``high-return college'' in addition to those that encourage ``high-return students'' to attend college.

\hfill

\end{abstract}

\clearpage

\section{Introduction}\label{Introduction}

Treatment variables of social and empirical interest often come in multiple varieties. 
For instance, years of education encompasses educational time investments across subjects and levels of schooling,
years of work experience encompasses time investments across a variety of tasks which develop different skills, and
healthcare expenditures encompass a wide variety of medical services.
When these differing components of composite treatments have different treatment effects, models which describe single effects of composite treatments on outcomes are misspecified, and the stable unit treatment value assumption (SUTVA) of \cite{rubin1974estimating} is violated.
This paper describes conditions under which standard instrumental variables (IV) methods identify convex weighted averages of unobserved treatments' effects, describes an alternative factor-augmented treatment effects model that allows IVs to place negative weights on unobserved treatments, and applies the new method to estimation of returns to college using the NLSY79.

For some applications involving treatments with unobserved treatment heterogeneity, estimated composite treatment effects may equally weight all component treatments, providing an estimate of the average effect of component treatments. In other cases, the estimated composite treatment effect may be inherited from a single unrepresentative component, which may not even share the sign of the average effect. Finally, if the variation in treatment used to estimate effects shifts individuals between unobserved component treatments sufficiently often, estimated composite treatment effects will reflect differences in effects between component treatments rather than a positively-weighted average effect of component treatments relative to nontreatment.

I begin with a theoretical discussion of standard instrumental variables methods when treatments come in multiple unobserved varieties.
I show that instruments which shift individuals into (but never out of or between) multiple versions of treatment identify convex weighted averages of unobserved treatments' effects, essentially extending the insights of \cite{imbens1994identification} and \cite{heckman2018unordered} to the setting with unobserved treatment heterogeneity.
I show similar results for the case in which instruments shift relatively few individuals with non-extreme treatment effects between treatment varieties, extending the insights of \cite{de2017tolerating} to the same setting.

In addition to results on standard methods, I also develop a novel factor-augmented treatment effects (FATE) model that defines treatment effects in terms of a small number (such as one) of particularly well-behaved instruments, while allowing other instruments to contribute to estimation.
The method estimates a convex average of effects of unmodeled component treatments without requiring observation of the component treatments.
The key insight motivating the method is that the varieties of treatment that individuals are shifted into by different instruments are outcome invariant, while the effects of unobserved varieties of treatment on outcomes are instrument invariant.
This suggests a factor structure for just-identified IV treatment effect estimands wherein IVs work through a lower dimensional number of unobserved treatments to affect a large number of outcomes in logically consistent ways.
When instruments are assumed to affect outcomes through a single treatment variety, this model is identical to that of 2SLS with multiple instruments, whereas it is identifical to just-identified IV for multiple instruments when the number of treatment varieties is assumed to be equal to the number of instruments.

I use the FATE model to estimate returns to college over the lifecycle using the NLSY79.
I find that college attendance associated with local labor market conditions produces a higher wage return profile than college attendance associated with local college proximity. 
Because local wage variation likely affects college attendance by making the outside option more or less attractive, it is reasonable to think that it does not shift individuals between versions of treatment, suggesting these (high) return estimates are more likely to be valid.
It follows that returns to college in the literature that use cost-shifters, which may shift individuals into low effort college because of income effects, may negatively weight high-return types of college and understate true returns to college on average.
Comparing the estimated factor-augmented treatment effects to those of just-identified IV shows reductions in standard errors, consistent with the intuition of FATE allowing multiple instruments to work together to estimate treatment effects (as in the extreme case of multiple-IV 2SLS).

Finally, I contribute to the literature on economic applications for factor models. Much of the existing work along these lines uses factor models to extract particularly relevant variation in data with multiple outcomes, where the multiple outcomes often include auxiliary measures such as test scores \citep{hansen2004effect, cunha2005separating, heckman2006effects, jiang2019women} or repeated outcomes of interest in panel data \citep{bai2009panel}. In these cases, the main use of factor models is to extract common sources of variation across outcomes to define otherwise unobserved types of individual-level heterogeneity that vary in their effects on outcomes. These factors act as controls in equations of interest to address omitted variable bias and enable identification of treatment effects of other observed variables. 
My factor model leans on this past work in the estimation procedure, but differs substantially in intuition. Rather than extracting particularly meaningful variation in data to identify factors that act as control variables, my factor model extracts particularly meaningful variation in estimated marginal effects to determine their structural interpretation.

The plan of the paper is as follows. Section 2 describes identification of treatment effects for standard IV methods and introduces a net uniform unordered monotonicity condition under which the IV estimand is a positively-weighted average of component treatment effects. Section 3 describes the factor-augmented treatment effects model. Section 4 describes the data for the application on returns to college. Section 5 discusses model estimates from the factor-augmented treatment effects model and their interpretation and relationship to those of multiple instrument 2SLS and just-identified IV. Section 6 concludes.

 %
%
%
%
\section{IV Estimands under Unobserved Treatment Heterogeneity}
In this section, I will describe conditions under which standard IV methods identify meaningful average effects of treatments with unobserved heterogeneity. 
In the context of the \cite{rubin1974estimating} causal model (RCM), unobserved treatment heterogeneity constitutes a violation of the Stable Unit Treatment Value Assumption, which requires there to be only one version of each treatment. 
I will start by presenting the problem using RCM potential outcome language and notation and going through some assumptions that will be maintained in subsequent sections.
Next, I will present sufficient conditions for treatment effect identification in the case where treatments are heterogeneous and their effects are homogeneous, wherein unobserved defier and complier treatments will play similar roles to those of defier and complier individuals in the more familiar case of homogeneous treatments with heterogeneous effects.
Afterward, I will allow for both unobserved treatment heterogeneity and treatment effect heterogeneity with a strict monotonicity assumption similar to those of \cite{imbens1994identification} and \cite{heckman2018unordered}.
I close the section by relaxing this monotonicity assumption for cases when treatments and their effects are heterogeneous by applying the insights of \cite{de2017tolerating} to allow for compliers (for each version of treatment) to cancel out defiers (for the same version of the treatment) who share their treatment effects.

\subsection{Potential Outcomes with Unobserved Treatment Heterogeneity}
I  work within a version of the \cite{rubin1974estimating,rubin1990comment} causal model similar to that used by \cite{imbens1994identification}, with reference as well to \cite{heckman1990varieties}, \cite{angrist1991sources}, and \cite{angrist1996identification}. 
The main innovation of this paper concerns an observed composite treatment for individual $i$, $D_i$, that is composed of $L$ unobserved mutually exclusive component treatments such that $D_i=\sum_{\ell=1}^LD_{i\ell}$, with the vector $\mathbf{D}_i = (D_{i0},D_{i1},...,D_{iL})$ denoting $i$'s treatment status for each version of the treatment, with $D_{i0}$ denoting the untreated state.
Because component treatments are binary and mutually exclusive, it follows that $D_i$ is binary as well.

I assume that the outcome $Y_i$ is observed with $Y_i = \sum_{\ell=0}^L D_{i\ell}Y_i(\ell)$, where $Y_i(\ell)$ gives the value of $Y_i(\mathbf{D}_i)$ when $D_{i\ell}=1$ and $D_{im}=0$ for all $m\neq \ell$. 
We  observe $Y_i(0)$ for individuals that do not select treatment, but not those that do, as is standard in the potential outcomes framework. However, with unobserved treatment heterogeneity, $D_i=1$ is not sufficient to determine the value of $D_{i\ell}$, other than to bound $\ell$ at $\ell>0$. It follows that not only do we not observe $Y_i(\ell)|D_i=0$ for any $i$ or $\ell>0$, but we also do not observe $Y_i(\ell)|D_{im}=1$ for any $i$ for $\ell \neq m$.%
\footnote{More precisely, we do observe $Y_i(\ell)$ for individuals who choose component treatment $\ell$, but I do not know the $\ell$ for which we are observing $Y_i(\ell)$.}
This muddling of potential outcomes for treated individuals is the central empirical challenge I address in this paper. Not only is the single composite treatment effect not observed for any individuals, it not well-defined. We instead are left with multiple component treatment effects, $Y_i(\ell)-Y_i(0)$, which can be differenced to define unobserved treatment switching effects, $Y_i(\ell)-Y_i(\ell')$ for all $\ell \neq \ell '$.

In the language of the Rubin causal model, unobserved treatment heterogeneity constitutes a violation of the stable unit treatment value assumption (SUTVA). In randomized controlled trials, random assignment to treatment ensures that $\mathbb{E}[Y_i(\ell)|D_{i\ell}=1] = \mathbb{E}[Y_i(\ell)|D_{im}=1]$ for all $\ell$ and $m$, including the cases where $\ell=0$ or $m=0$. Experimental control efforts are intended to ensure that $D_i=D_{i\ell}$ for a given $\ell$ for all individuals in an observed treatment arm. In other words, the administration of the treatment is explicitly designed to hold equal between treated individuals all variations that are expected to substantively alter the nature of the treatment.\footnote{Even well-designed experiments differ in this respect. Variation in subtle priming from the outward affect or word choices of research staff administering a medical treatment might not complicate treatment effect identification for treatments of physical ailments but may do so for treatments of psychological ailments.} The control of the trial ensures identification of treatment effects that are entirely driven by a single component treatment, while the randomization ensures that the effects can be estimated via comparisons of treated and untreated individuals.

In observational studies such as the application below on the wage returns to college, treatment assignments are not explicitly randomized, individuals who select treatment are not necessarily representative of the population, and versions of treatment received by individuals are not controlled.
To address the lack of randomization of treatment, it is common for researchers to make use of instrumental variables, variables which are assumed to be as-good-as-random that shift individuals into (or out of) treatment without otherwise affecting outcomes.
The implications for treatment effect estimation of selected subpopulations of individuals responding to instrumental variables are well-understood and commonly addressed in empirical work.
The implications of individuals responding to instrumental variables by selecting particular unmodeled varieties of treatment are less well-understood, but produce similar problems, as I will show.
This is especially of interest for estimates of returns to college, considering the large literature detailing differences in effects of college on outcomes depending on variations in treatment such as the subjects of study \citep{hastings2013some,kirkeboen2016field} or the quality of the college \citep{black2006estimating,dillon2020consequences}.

In addition to observing the outcome and composite treatment, I assume that an instrumental variable $Z_i \in \mathcal{Z}$ is observed as well, composite and component treatments $D_i(z)$ and $D_{i\ell}(z)$ representing random variables that give the value of each respective treatment if $Z_i=z$. I assume that $(Y_i,D_i,Z_i)$ is observed for a random sample of the population. I assume that instrumental variables satisfy several conditions, which I will detail.
\begin{condition}[Independence]\label{condition:independence}
Let $Z$ be a random variable such that $D_{i\ell}(z)$ is independent of $Z_i$ for all $\ell=0,1,...,L$ and all $z \in \mathcal{Z}$.
\end{condition}
\noindent Condition \ref{condition:independence} would be sufficient to identify the effect of $Z_i$ on $D_{i\ell}$ if $D_{i\ell}$ were observed. It follows from this condition and the definition of $D_i$ that $D_i$ is independent of $Z_i$, further implying that the effect of $Z_i$ on the composite treatment, $D_i$, is identified as well.

In addition to the multiple treatments version of independence just given, I also present a multiple treatments version of exclusion.
\begin{condition}[Exclusion]\label{condition:exclusion}
Let $Z$ be a random variable such that $Y(\ell)$ is independent of $Z_i$ for all $\ell=0,1,...,L$ and all $z \in \mathcal{Z}$.
\end{condition}
\noindent The standard composite treatment version of exclusion is likely to be violated in settings with SUTVA violations, while the multiple treatment version allows will not be for well-chosen instruments.%
\footnote{\cite{angrist1996identification} describe SUTVA as as ruling out treatment spillovers, while modeling outcomes, treatment assignments, and instruments for all individuals as vectors. The arguably simpler presentation in \cite{imbens1994identification} works, like this paper, at the individual level and makes no mention of SUTVA, implicitly ruling out problematic spillovers and other forms of unobserved treatment heterogeneity with their independence assumption (which includes what I and others call exclusion).}
This is readily seen in the regression-based treatment effects model with homogeneous effect $\lambda_\ell$ for treatment $\ell$, \hfill
\begin{equation}
    \begin{split}
    Y_i &= \lambda_0+\sum_{\ell=1}^LD_{i\ell}(Z_i)\lambda_\ell+\epsilon_i \\
    &= \lambda_0+0D_i(Z_i)+(\epsilon_i+\sum_{\ell=1}^LD_{i\ell}(Z_i)\lambda_\ell),
    \end{split}
\end{equation}
where the term $\sum_{\ell=1}^LD_{i\ell}(Z_i)\lambda_\ell$ is in the error term in the homogeneous treatment version of the model in the second line. The $Y_i(1)$ outcome is clearly dependent on $Z_i$ via channels other than $D_i$, unless the instrument only ever induces a particular type of treatment (such as in the case of a controlled experiment, or an intent-to-treat instrument in the same context). 

Conditions \ref{condition:independence} and \ref{condition:exclusion} imply that the effect of $Z_i$ on $Y_i$ is identified in addition to the effect of $Z_i$ on $D_i$. The effect of $Z_i$ on $Y_i$, defined for a comparison of two points $(z,z')$ in the support of $\mathcal{Z}$, is given by \hfill
\begin{equation}\label{Reduced_Form}
\begin{split}
    \mathbb{E}[Y_i&|Z_i=z] - \mathbb{E}[Y_i|Z_i=z'] \\
    =
    &\mathbb{E}[\sum_{\ell=1}^L D_{i\ell}(z)Y_i(\ell)+(1-\sum_{\ell=1}^L D_{i\ell}(z))Y_i(0)|Z_i=z] \\
    -&\mathbb{E}[\sum_{\ell=1}^L D_{i\ell}(z')Y_i(\ell)+(1-\sum_{\ell=1}^L D_{i\ell}(z'))Y_i(0)|Z_i=z'].
    \end{split}
\end{equation}
Conditions \ref{condition:independence} and \ref{condition:exclusion} imply that this is equal to
\begin{equation}\label{Nonparametic_Reduced_Form_Decomposition}
\begin{split}
    \sum_{\ell=1}^L
    \mathbb{E}
    &\Big[(D_{i\ell}(z)-D_{i\ell}(z'))(Y_i(\ell)-Y_i(0)) \Big] \\
    = 
    &\sum_{\ell=1}^L \Pr[D_{i\ell}(z)-D_{i\ell}(z')=1]\mathbb{E}[Y_i(\ell)-Y_i(0)|D_{i\ell}(z)-D_{i\ell}(z')=1] \\
    - 
    &\sum_{\ell=1}^L \Pr[D_{i\ell}(z)-D_{i\ell}(z')=-1]\mathbb{E}[Y_i(\ell)-Y_i(0)|D_{i\ell}(z)-D_{i\ell}(z')=-1].
    %
    \end{split}
\end{equation}
The IV estimand for the effect of $D_i$ on $Y_i$ for instrument comparison $z,z'$ is obtained by dividing the average effect of $Z_i$ on $Y_i$ by the average effect of $Z_i$ on $D_i$, 
\begin{equation}\label{IV_estimand}
    \pi_{z,z'} =
    \frac{
    \mathbb{E}[Y_i|Z_i=z] - \mathbb{E}[Y_i|Z_i=z']
    }
    {
    P(z)-P(z')
    },
\end{equation}
where $P(z) = \mathbb{E}[D_i|Z_i=z]$ gives the probability of treatment for instrument value $z \in \mathcal{Z}$. There are three potential problems with this estimand even when conditions \ref{condition:independence} and \ref{condition:exclusion} are satisfied.  
In brief, problems arise with this estimand if instruments do not appreciably affect composite treatment status, if instruments shift some individuals into treatments and some out, or if instruments shift individuals into some component treatments and out of others.

First, this estimand is indeterminant or undefined if $P(z)-P(z') = 0$. This is addressed with the usual testable condition that the instrument has an appreciable effect on observed composite treatment status,
\begin{condition}[Relevance]\label{condition:relevance}
Let $Z$ be a random variable such that $P(z)$ is a nontrivial function of $z$.
\end{condition}
\noindent It follows from condition \ref{condition:relevance} that the instrument must also have an effect on at least one component treatment, such that the unobserved value $P_\ell(z) = \mathbb{E}[D_{i\ell}|Z_i=z]$ is also a nontrivial function of $z$ for some value of $\ell$.

The second and third potential problems with the estimand in equation (\ref{IV_estimand}) involve the negative second term. 
The case in which instruments shift some individuals into a homogeneous treatment and some individuals out, with the treatment effects of those in the minority (defier individuals) being negatively weighted is addressed by the monotonicity condition of \cite{imbens1994identification}, which constrains this term to zero by assuming all individuals respond to the instrument in the same way.
The closely-related problem presented by unobserved treatment heterogeneity is that instruments might shift individuals into some versions of treatment and out of others, with the treatment effects of treatments in the minority (defier treatments) being negatively weighted.
Multiple sufficient conditions of varying plausibility in applications address this second concern.
I will begin with less plausible (but simpler) sufficient conditions and work toward more plausible ones.

\subsection{Heterogeneous Treatments with Homogeneous Effects}
The simplest ways of addressing the problematic second term of equation (\ref{IV_estimand}) make assumptions that imply that $\sum_{\ell=1}^L \Pr[D_{i\ell}(z)-D_{i\ell}(z')=-1] = 0$ for all $\ell$ and all $i$ (assuming without loss of generality that $P(z)>P(z')$).
An example of this approach is in \cite{imbens1994identification}, who implicitly assume $L=1$, and that instruments only shift individuals into (not out of) treatment.
I begin by considering conditions that mirror this approach in the setting of unobserved treatment heterogeneity. First, I consider the case where each unobserved type of treatment has homogeneous treatment effects.
\begin{condition}[Homogeneous Treatment Effects]
\label{condition:homogeneous_treatment_effects}
For all $\ell=1,2,...,L$, define $\lambda_\ell$ as a constant that satisfies $\lambda_\ell=Y_i(\ell)-Y_i(0)$.
\end{condition}
\noindent Condition \ref{condition:homogeneous_treatment_effects}, combined with \ref{condition:independence} and \ref{condition:exclusion}, implies that the effect of $Z_i$ on $Y_i$ from equations (\ref{Reduced_Form}) and (\ref{Nonparametic_Reduced_Form_Decomposition}) can be rewritten as 
\begin{equation}\label{Homogeneous_Treatments_Decomposition}
\begin{split}
    \mathbb{E}[Y_i|Z_i=z] - \mathbb{E}[Y_i|Z_i=z']
    = 
    &\sum_{\ell=1}^L(P_{\ell}(z)-P_{\ell}(z'))\lambda_\ell \\
     =
     & \sum_{\ell=1}^L I(P_\ell(z) \geq P_\ell(z'))(P_{\ell}(z)-P_{\ell}(z'))\lambda_\ell \\
     -
     & \sum_{\ell=1}^L I(P_\ell(z) \leq P_\ell(z'))(P_{\ell}(z')-P_{\ell}(z))\lambda_\ell,
    \end{split}
\end{equation}
where $I(\cdot)$ is an indicator function that takes a value of unity if its argument is true and zero otherwise.

Even with the strong assumption of homogeneous treatment effects, the IV estimand in equation (\ref{IV_estimand}) can be uninformative about treatment effects.
To see this, consider the example with high-return college and low-return college as the unobserved versions of treatment, with a randomly assigned low-return-major contingent college scholarship used as an instrument for college attendance.
If returns to high-return college are double those of low-return college and this instrument shifts $x$ individuals from non-college into low-return college and $x$ individuals from high-return college into low-return college, the IV estimand in (\ref{IV_estimand}) will fail to provide information regarding returns to any type of college.
In the model, this setting constitutes $L=2$ with $\lambda_2 = 2\lambda_1$ and $P_1(z)-P_1(z') = -2(P_2(z)-P_2(z'))$, which produces a value of zero on the right hand side of equation (\ref{Homogeneous_Treatments_Decomposition}).

To avoid some versions of treatment having their effects negatively weighted in the IV estimand, additional assumptions are required even for with treatments with constant effects and instruments that satisfy independence, exclusion, and relevance. A simple sufficient condition imposes that that instrument never induces changes in unobservable intensive margin treatment status.
\begin{condition}[Uniform Treatment Responses]\label{condition:uniform_treatment_responses}
For all $z,z'\in \mathcal{Z}$, either $P_{\ell}(z)\geq P_{\ell}(z')$ for all $\ell$, or $P_{\ell}(z)\leq P_{\ell}(z')$ for all $\ell$.
\end{condition}
\noindent Condition \ref{condition:uniform_treatment_responses} does for treatments with heterogeneous effects what the monotonicity condition of \cite{imbens1994identification} does for individuals with heterogeneous treatment effects.

If the above conditions hold, the estimand in equation (\ref{IV_estimand}) has a causal interpretation.
\begin{thm}\label{theorem:UTR_effect}
    If Conditions 1-5 hold, the following weighted average component treatment effect is identified: \hfill
    \begin{equation*}
    \pi_{z,z'} =
    \sum_{\ell=1}^L 
    \frac{
    P_{\ell}(z)-P_{\ell}(z')
    }
    {
    P(z)-P(z')
    }
    \lambda_\ell,
\end{equation*}
with $(P_{\ell}(z)-P_{\ell}(z'))/(P(z)-P(z')) \geq 0$ for all $\ell$.
\end{thm}
\begin{proof}
Let condition \ref{condition:uniform_treatment_responses} hold with $P_\ell(z) \geq P_\ell(z')$ for all $\ell$. 
It immediately follows that $(P_{\ell}(z)-P_{\ell}(z'))/(P(z)-P(z')) \geq 0$ for all $\ell$. 
It also follows that the final term in equation (\ref{Homogeneous_Treatments_Decomposition}) is zero, and \hfill
\begin{equation*}
    \mathbb{E}[Y_i|Z_i=z] - \mathbb{E}[Y_i|Z_i=z']
    = 
    \sum_{\ell=1}^L 
    (P_{\ell}(z)-P_{\ell}(z'))\lambda_\ell.
\end{equation*}
Diving both sides of (\ref{Homogeneous_Treatments_Decomposition}) by $(P(z)-P(z'))$ shows that the weighted average component treatment effect above can be obtained from moments of the joint distribution of $(Y_i,D_i,Z_i)$.
\end{proof}

An economic model of the college attendance decision is helpful for understanding the connection between unobserved treatment heterogeneity and treatment effect heterogeneity. 
For clarity, consider the case of two unobserved component treatments with one observed composite treatment, such that $L=2$.
The discrete choice model is the same as the case where all treatments are observed, such as in \cite{heckman2018unordered} and \cite{mountjoy2022community}.
Formally, I define individual $i$'s utility from treatment decision $D_{i\ell}(Z_i)$ as  $D^*_{i\ell}(Z_i)$, defined as
\begin{equation}
    \begin{split}
        D^*_{i0}(Z_i) &= 0 \\
        D^*_{i1}(Z_i) &= U_{i1}+\mu_1(Z_i) \\
        D^*_{i2}(Z_i) &= U_{i2}+\mu_2(Z_i),
    \end{split}
\end{equation}
where the utility from noncollege is normalized to zero, and individuals choose the treatment status that gives them the highest utility, such that
\begin{equation}\label{discrete_choice_model}
    \begin{split}
        D_{i0}(Z_i) &= I(D^*_{i1}(Z_i)<0, D^*_{i2}(Z_i)<0) \\
        D_{i1}(Z_i) &= I(D^*_{i1}(Z_i)\geq D^*_{i2}(Z_i),D^*_{i1}(Z_i)\geq 0) \\
        D_{i2}(Z_i) &= I(D^*_{i2}(Z_i)\geq D^*_{i1}(Z_i), D^*_{i2}(Z_i)\geq 0).
    \end{split}
\end{equation}

The assumption that all individuals are uniformly affected by instruments ($\mu_i(Z_i) = \mu(Z_i)$ for all $i$) is unnecessary for the methods in this paper, but permits a helpful graphical representation of the model. 
Figure \ref{fig:Net_monotonicity_illustration} represents the college attendance decision model in (\ref{discrete_choice_model}) with $(U_{1i},U_{2,i})$ drawn from a distribution that implies that each point enclosed by the ellipse contains a common population density. 
The shares of the population attending each type of college at different values of $Z$ are given by the area inside of the ellipse representing the joint distribution of $(D_1^*(Z),D_2^*(Z))$. 
The three panels each consider an instrument comparison relative to a common baseline with a relatively low college attendance rate.
The instrument comparisons in Panels B and C satisfy condition \ref{condition:uniform_treatment_responses}, while the comparison in Panel A does not.

To fix ideas, consider that Panel A shows a control group, individuals for whom $Z_i=z$, and a treated group who are randomly selected to receive a STEM-major-contingent college scholarship, for whom $Z_i=z'$.
The STEM-contingent scholarship induces individuals who would not otherwise attend college at all into STEM college ($\ell=1$), but also induces individuals who would otherwise attend college with a non-STEM major into STEM college.
In this case, $P_1(z')>>P_1(z)$ with $P_2(z')<P_2(z)$, violating condition \ref{condition:uniform_treatment_responses}.
If unobserved treatment $D_1$ has larger (smaller) treatment effects than treatment $D_2$, estimating effects of college using the comparison $(z,z')$ will be at risk of overstating (understating) the true returns of any type of college.
Panels B and C display individuals in alternative treatment arms for whom $Z_i=z''$ and $Z_i=z'''$ which represent receipt of other randomly-assigned incentives for college that have less-extreme compositional effects on the shares of individuals attending each type of college.
\begin{SidewaysFigure}

    \centering
    \begin{subfigure}[t]{0.4\textwidth}
    \raggedright
        Panel A
        \begin{center}
        \resizebox{\textwidth}{!}{
\begin{tikzpicture}
\draw [-{Latex[round]}] (-4,-3.5) -- (5,-3.5);
\draw [-{Latex[round]}] (-4,-3.5) -- (-4,3);
\node[label=below:{$D_1^*$} ]   () at (4.5, -3.5) {};
\node[label=below:{$D_2^*$} ]   () at (-4.5, 2.9) {};
\node[label=below:{$D=0$} ]   () at (-2.5, -1.5) {};
\node[label=above:{$D_2=1$} ]   () at (-2.5, 1.5) {};
\node[label=below:{$D_{1}=1$} ]   () at (2.5, -1.5) {};
\node[label=above:z]   () at (.8, -1.1) {};
\node[label=above:z']   () at (1.8, -1.1) {};

\draw [-] (0,0) -- (3,3);
\draw [-] (0,0) -- (0,-3.5);
\draw [-] (0,0) -- (-4,0);
\draw[rotate=0][gray, dashed] (-1.5,-0.5) ellipse (60pt and 30pt);
\draw[rotate=0] (-0.5,-0.5) ellipse (60pt and 30pt);
\draw [-{Latex[round]}] (0.6,-0.5) -- (1.6,-0.5);
\end{tikzpicture}
}
\end{center}
\end{subfigure}
\hspace{0.05\textwidth}%
    \begin{subfigure}[t]{0.4\textwidth}
    \raggedright
        Panel B
        \begin{center}
        \resizebox{\textwidth}{!}{
\begin{tikzpicture}
\draw [-{Latex[round]}] (-4,-3.5) -- (5,-3.5);
\draw [-{Latex[round]}] (-4,-3.5) -- (-4,3);
\node[label=below:{$D_1^*$} ]   () at (4.5, -3.5) {};
\node[label=below:{$D_2^*$} ]   () at (-4.5, 2.9) {};
\node[label=below:{$D=0$} ]   () at (-2.5, -1.5) {};
\node[label=above:{$D_2=1$} ]   () at (-2.5, 1.5) {};
\node[label=below:{$D_{1}=1$} ]   () at (2.5, -1.5) {};
\node[label=above:z]   () at (.8, -1.1) {};
\node[label=above:z'']   () at (1.85, -0.8) {};

\draw [-] (0,0) -- (3,3);
\draw [-] (0,0) -- (0,-3.5);
\draw [-] (0,0) -- (-4,0);
\draw[rotate=0][gray, dashed] (-1.5,-0.5) ellipse (60pt and 30pt);
\draw[rotate=0] (-0.5,-0.25) ellipse (60pt and 30pt);
\draw [-{Latex[round]}] (0.6,-0.5) -- (1.6,-0.2);
\end{tikzpicture}
}
\end{center}
\end{subfigure}
\hspace{0.05\textwidth}%
    \begin{subfigure}[t]{0.4\textwidth}
    \raggedright
        Panel C
        \begin{center}
        \resizebox{\textwidth}{!}{
\begin{tikzpicture}
\draw [-{Latex[round]}] (-4,-3.5) -- (5,-3.5);
\draw [-{Latex[round]}] (-4,-3.5) -- (-4,3);
\node[label=below:{$D_1^*$} ]   () at (4.5, -3.5) {};
\node[label=below:{$D_2^*$} ]   () at (-4.5, 2.9) {};
\node[label=below:{$D=0$} ]   () at (-2.5, -1.5) {};
\node[label=above:{$D_2=1$} ]   () at (-2.5, 1.5) {};
\node[label=below:{$D_{1}=1$} ]   () at (2.5, -1.5) {};
\node[label=above:z]   () at (-.05, 0.1) {};
\node[label=above:z''']   () at (0.75, 0.9) {};

\draw [-] (0,0) -- (3,3);
\draw [-] (0,0) -- (0,-3.5);
\draw [-] (0,0) -- (-4,0);
\draw[rotate=0][gray, dashed] (-1.5,-0.5) ellipse (60pt and 30pt);
\draw[rotate=0] (-0.7,0.25) ellipse (60pt and 30pt);
\draw [-{Latex[round]}] (-0.4,0.4) -- (0.3,1.1);
\end{tikzpicture}
}
\end{center}
\end{subfigure}
\caption{Instrument Comparisons with Two Unobserved Component Treatments}
\label{fig:Net_monotonicity_illustration}
   \begin{minipage}{1\linewidth}
\smallskip
\footnotesize
\emph{Notes:} Graphic representations of instrument comparisons that satisfy independence, exclusion, and relevance in the context of decision model (\ref{discrete_choice_model}). 
Panel A shows an instrument comparison that satisfies Heckman and Pinto's \citeyearpar{heckman2018unordered} unordered monotonicity condition, but violates uniform unordered monotonicity and net uniform unordered monotonicity (there are more type 2 defiers than compliers).
Panel B shows an instrument comparison that violates uniform unordered monotonicity, but may satisfy net uniform unordered monotonicity with low treatment effect heterogeneity.
Panel C's instrument comparison satisfies uniform unordered monotonicity and net uniform unordered monotonicity.
\end{minipage}
\end{SidewaysFigure}

\subsection{Heterogeneous Treatments with Heterogeneous Effects}
There is a conceptual point to be made that a fair amount of variation in effects of observed composite treatments between individuals may be attributable to variation in versions of treatment chosen, rendering condition \ref{condition:homogeneous_treatment_effects} less restrictive than it is in the usual homogeneous treatments case.
For instance, continuing with the STEM/non-STEM example above, consider the individuals who are at the margin of non-college and STEM college, who are extremely far from all other margins. 
Such individuals will generally respond to instruments (if they respond) by attending a particular type of college, so there is no practical distinction for them between treatment heterogeneity and treatment effect heterogeneity --- they are, in the running example, STEM-margin individuals at the margin of STEM college.
However, individuals who are close to the line $D_1^*=D_2^*$ are at the intensive margin between types of college, such that their returns to college cannot be completely described by attributing to them individual-specific returns; they may choose either type of college and will have different outcomes depending on their choice.%
\footnote{For instance, in Figure \ref{fig:Net_monotonicity_illustration} many individuals along the top of the $D=0$ partition when $Z_i=z$ choose treatment 1 when $Z_i=z'$ (Panel A), but they choose treatment 2 when $Z_i=z'''$ (Panel C).}
It follows that while condition \ref{condition:homogeneous_treatment_effects} is less restrictive than the corresponding assumption with homogeneous treatments, it nonetheless may fail to capture meaningful variation in treatment effects between populations.

This section replaces conditions \ref{condition:homogeneous_treatment_effects} and \ref{condition:uniform_treatment_responses} with an alternative condition that allows for treatment effect identification when unobserved component treatments have individual-level treatment effect heterogeneity. 
\begin{condition}[Uniform Unordered Monotonicity]\label{condition:uniform_unordered_monotonicity}
    For all $z,z' \in \mathcal{Z}$, either  
    $D_{i\ell}(z)\geq D_{i\ell}(z')$   for   all   $i=1,2,...,N$   and   all   $\ell=1,2,...L$,      or  
    $D_{i\ell}(z)\leq D_{i\ell}(z')$   for   all   $i=1,2,...,N$   and   all   $\ell=1,2,...L$.
\end{condition}
\noindent This condition strengthens Heckman and Pinto's \citeyearpar{heckman2018unordered} unordered monotonicity assumption to require that all treatments respond to instruments uniformly, that is, if an instrument makes one component treatment more likely, it must weakly make all others more likely as well. In related work on observed treatments, \cite{kline2016evaluating} describe a similar condition that restricts changes in the treatment decision to a single observed component treatment in the presence of multiple observed varieties of treatment, while \cite{hull2018isolateing} describes a condition that restricts changes in treatment decision from one of multiple observed untreated varieties to a single observed treatment variety.

Condition \ref{condition:uniform_unordered_monotonicity} implies that the IV estimand in (\ref{IV_estimand}) has a causal interpretation as a weighted average of effects of complier treatments on complier individuals.
\begin{thm}
    If Conditions 1-3 and 6 hold, the following component-weighted local average treatment effect is identified: \hfill
    \begin{equation*}
    \pi_{z,z'} =
    \sum_{\ell=1}^L 
    \frac{
    P_{\ell}(z)-P_{\ell}(z')
    }
    {
    P(z)-P(z')
    }
    \mathbb{E}[Y_i(\ell)-Y_i(0)|D_{i\ell}(z) \neq D_{i\ell}(z')],
\end{equation*}
with $(P_{\ell}(z)-P_{\ell}(z'))/(P(z)-P(z')) \geq 0$ for all $\ell$.
\end{thm}
\begin{proof}
Let Condition \ref{condition:uniform_unordered_monotonicity} hold with $D_{i\ell}(z) \geq D_{i\ell}(z')$ for all $i$ and all $\ell$. 
It immediately follows that $(P_{\ell}(z)-P_{\ell}(z'))/(P(z)-P(z')) \geq 0$ for all $\ell$. 
It also follows that the final term in equation (\ref{Homogeneous_Treatments_Decomposition}) is zero, and \hfill
\begin{equation*}
\begin{split}
    \mathbb{E}[Y_i|Z_i=z] - \mathbb{E}[Y_i|Z_i=z'] 
    = 
    \sum_{\ell=1}^L 
     (P_{\ell}(z)-P_{\ell}(z'))\mathbb{E}[Y_i(\ell)-Y_i(0)|D_{i\ell}(z)-D_{i\ell}(z')=1].
    \end{split}
\end{equation*}
Diving both sides of this by $(P(z)-P(z'))$ shows that the weighted average local average component treatment effect above can be obtained from moments of the joint distribution of $(Y_i,D_i,Z_i)$.
\end{proof}

Condition \ref{condition:uniform_unordered_monotonicity} is quite strict. In the context of the decision model in (\ref{discrete_choice_model}), it will be satisfied in only two narrow cases. The first is the case in which the support of $\mathcal{Z}$ contains only two values, $(z,z')$, where $P(z)=0$. It follows that any comparison $(z,z')$ trivially satisfies a wide range of monotonicity conditions, including Condition \ref{condition:uniform_unordered_monotonicity}.%
\footnote{This amounts in practice to using an indicator $I(P(Z_i)=0)$ as an instrument. The benefits of observing values of $Z_i$ at which no individuals receive treatment is discussed elsewhere for the homogeneous treatment setting, such as by \cite{heckman1990varieties}, \cite{angrist1991sources}, and \cite{imbens1994identification}. Treatment assignment with perfect compliance in the case of a randomized trial falls into this case, so such settings naturally satisfy uniform unordered monotonicity.}
The second is the case in which instruments make all treatments more or less attractive to all individuals by a constant amount. In Figure \ref{fig:Net_monotonicity_illustration}, this is illustrated in Panel C, where the slope of the effect of instruments on preferences for treatments is parallel to the unobserved decision margin, $D2^*=D_1^*$.%
\footnote{
Heckmand and Pinto's \citeyearpar{heckman2018unordered} unordered monotonicity condition for multiple observed treatments, in the context of model \ref{discrete_choice_model}, allows for the slope of the effect of instruments to be parallel to any decision margin, one example of which is shown in Panel A of Figure \ref{fig:Net_monotonicity_illustration}. Uniform unordered monotonicity thus allows for $1/(L+1)$ as many admissible slopes for the effects of instruments on treatment utilities.}

\subsection{Tolerating Defiance}
It is difficult to imagine an instrument, other than treatment assignment in a randomized trial with perfect compliance, that strictly satisfies Condition \ref{condition:uniform_unordered_monotonicity}. 
For instance, exogenous college cost shifters presumably have income effects, which may cause some individuals to switch to less-difficult versions of college, while college distance shifters may cause some individuals to switch from whatever variety (quality, etc) of college they otherwise would have attended to the nearby college.
As in the case of homogeneous treatment monotonicity, it is intuitive that small violations are unlikely to be particularly harmful.
\cite{de2017tolerating} formalizes this intuition and provides a weaker alternative to Imbens and Angrists' \citeyearpar{imbens1994identification} monotonicity assumption in the setting with homogeneous treatments.
This section formalizes this intuition in the setting with heterogeneous treatments and provides a substantially weaker alternative to Condition \ref{condition:uniform_unordered_monotonicity} that is more likely to be satisfied in empirical applications.

I begin by defining component-treatment-specific versions of the compliers, defiers, always-takers, and never-takers described by \cite{angrist1996identification} and others.
Considering instrument comparison $(z,z')$, individual $i$ is an $\ell$ complier with if $D_{i\ell}(z)=1$ and $D_{i\ell}(z')=0$, an $\ell$ defier if $D_{i\ell}(z)=0$ and $D_{i\ell}(z')=1$, an $\ell$ always-taker if $D_{i\ell}(z)=1$ and $D_{i\ell}(z')=1$, and an $\ell$ never-taker if $D_{i\ell}(z)=0$ and $D_{i\ell}(z')=0$.
Let $C_{i\ell}(z,z')=1$ if individual $i$ is an $\ell$ complier ($C_{i\ell}(z,z')=0$ otherwise), with the frequency of $\ell$ compliers given by $P_{c\ell}(z,z')$. 
Let $F_{i\ell}(z,z')=1$ if $i$ is an $\ell$ defier ($F_{i\ell}(z,z')=0$ otherwise), with the frequency of $\ell$ defiers given by $P_{f\ell}(z,z')$. 
Allowing for the presence of defiers and assuming that conditions 1-3 hold, the IV estimand in (\ref{IV_estimand}) is equivalent to
\begin{equation}\label{IV_estimand_Defiers}
    \pi_{z,z'} =
    \frac{
    \sum_{\ell=1}^L
    P_{c\ell}(z,z')\mathbb{E}[Y_i(\ell) - Y_i(0)|C_{i\ell}(z,z')=1]-P_{f\ell}(z,z')\mathbb{E}[Y_i(\ell) - Y_i(0)|F_{i\ell}(z,z')=1]
    }
    {
    \sum_{\ell=1}^L
    P_{c\ell}(z,z')-P_{f\ell}(z,z')
    }.
\end{equation}
Condition \ref{condition:uniform_unordered_monotonicity} sets $P(F_\ell)=0$ in this equation, producing a weighted average across treatment varieties of complier treatment effects.

As an alternative to Condition \ref{condition:uniform_unordered_monotonicity}, I propose a condition in which compliers exist for each component treatment that are able to cancel out the contributions of defiers to the IV estimand. 
\begin{condition}[Net Uniform Unordered Monotonicity]\label{condition:net_uniform_unordered_monotonicity}
    For all $z,z' \in \mathcal{Z}$, there exists a $P_{c\ell}^f(z,z')$ share of the population uniquely defined by $C^f_{i\ell}(z,z')=1$ such that $C^f_{i\ell}(z,z')=1 \implies C_{i\ell}(z,z')=1$, $P_{c\ell}^f(z,z')=P_{f\ell}(z,z')$, and $\mathbb{E}[Y_i(\ell) - Y_i(0)|C^f_{i\ell}(z,z')=1] = \mathbb{E}[Y_i(\ell) - Y_i(0)|F_{i\ell}(z,z')=1]$ for all $\ell$.
\end{condition}
\noindent Condition \ref{condition:net_uniform_unordered_monotonicity} is effectively similar to Conditions \ref{condition:homogeneous_treatment_effects} and \ref{condition:uniform_treatment_responses} holding for many unobserved subgroups of the population, rather than for the population as a whole.
It allows for some individuals to respond to instruments by switching out of component treatment $\ell$ as long as other individuals switch in to component treatment $\ell$ who happen to have overlapping treatment effects. Situations such as those pictured in Panel B of Figure \ref{fig:Net_monotonicity_illustration} conceivably satisfy Condition \ref{condition:net_uniform_unordered_monotonicity} if treatment effects are nearly homogeneous, while instruments that affect treatment preferences in ways that narrowly deviate from that pictured in Panel C likely satisfy this condition even with extreme treatment effect heterogeneity. 

To establish the causal interpretation of the IV estimand under Condition \ref{condition:net_uniform_unordered_monotonicity}, I define treatment $\ell$ surviving compliers as individuals for whom $C_{i\ell}^s(z,z')=1$ where $C_{i\ell}^s(z,z') = I(C_{i\ell}(z,z')=1,C^f_{i\ell}(z,z')\neq 1)$, who are present in the population with frequency $P_{c\ell}^s(z,z') = Pr(C_{i\ell}^s(z,z')=1)$. The following theorem defines a the IV estimand as a positively-weighted average of a subset of treatments for a subset of individuals.
\begin{thm}\label{survivor_complier_IV_estimand}
    If Conditions 1-3 and \ref{condition:net_uniform_unordered_monotonicity} hold, the following component-weighted local average treatment effect is identified: \hfill
    \begin{equation}\label{nuum_estimand}
    \pi_{z,z'} =
    \sum_{\ell=1}^L 
    \frac{
    P_{c\ell}^s(z,z')
    }
    {
    P_{c}^s(z,z')
    }
    \mathbb{E}[Y_i(\ell)-Y_i(0)|C_{i\ell}^s(z,z')=1],
\end{equation}
with $P_{c\ell}^s(z,z')/P_{c}^s(z,z') \geq 0$ for all $\ell$, where $P_{c}^s(z,z') = \sum_{\ell=1}^LP_{c\ell}^s(z,z')$.
\end{thm}
\begin{proof}
By the definition of $P_{c}^s(z,z')$, $P_{c\ell}^s(z,z')/P_{c}^s(z,z') \geq 0$ if $P_{c\ell}^s(z,z') \geq 0$ for all $\ell$, which holds because $P_{c\ell}^s(z,z')$ is a probability bounded by $[0,1]$.
Under conditions 1-3 and \ref{condition:net_uniform_unordered_monotonicity}, the reduced form effect of $Z_i$ on $Y_i$ evaluated at comparison $(z,z')$ given in (\ref{Reduced_Form}) and (\ref{Nonparametic_Reduced_Form_Decomposition}) can be expanded as
\begin{equation*}
\begin{split}
   &\mathbb{E}[Y_i|Z_i=z] - \mathbb{E}[Y_i|Z_i=z']
   =
    \sum_{\ell=1}^L P^s_{c\ell}(z,z')\mathbb{E}[Y_i(\ell) - Y_i(0)|C^s_{i\ell}(z,z')=1] \\
    +&P^f_{c\ell}(z,z')\mathbb{E}[Y_i(\ell) - Y_i(0)|C^f_{i\ell}(z,z')=1]
    -P_{f\ell}(z,z')\mathbb{E}[Y_i(\ell) - Y_i(0)|F_{i\ell}(z,z')=1],
    \end{split}
\end{equation*}
which reduces to 
\begin{equation*}
\begin{split}
   &\mathbb{E}[Y_i|Z_i=z] - \mathbb{E}[Y_i|Z_i=z']
   =
    \sum_{\ell=1}^L P^s_{c\ell}(z,z')\mathbb{E}[Y_i(\ell) - Y_i(0)|C^s_{i\ell}(z,z')=1]
    \end{split}
\end{equation*}
because Condition \ref{condition:net_uniform_unordered_monotonicity} requires that $P_{c\ell}^f(z,z')=P_{f\ell}(z,z')$ and $\mathbb{E}[Y_i(\ell) - Y_i(0)|C^f_{i\ell}(z,z')=1] = \mathbb{E}[Y_i(\ell) - Y_i(0)|F_{i\ell}(z,z')=1]$ for all $\ell$. Dividing both sides by $P(z)-P(z')$ yields the IV estimand, where $P(z)-P(z') = \sum_{\ell=1}^L
    P_{c\ell}(z,z')-P_{f\ell}(z,z')$ under Conditions 1-3, as shown in equation (\ref{IV_estimand_Defiers}). Condition \ref{condition:net_uniform_unordered_monotonicity} implies that $\sum_{\ell=1}^L
    P_{c\ell}(z,z')-P_{f\ell}(z,z') = \sum_{\ell=1}^L
    P_{c\ell}^s(z,z')+P_{c\ell}^f(z,z')-P_{f\ell}(z,z')$, with $P_{c\ell}^f(z,z')=P_{f\ell}(z,z')$, such that $P(z)-P(z') = P_{c\ell}^s(z,z')$.
It follows that the weighted average of surviver-compliers' component treatment effects given by theorem  \ref{survivor_complier_IV_estimand} is obtained from the joint distribution of $(Y_i,D_i,Z_i)$.
\end{proof}

Several alternative assumptions are possible, which I leave to future study. For instance, using multiple instrument comparisons $(z,z')$ and $(z'',z''')$ allows for the possibility that defiers for one comparison are compliers for the other, which is sufficient for identification if the well-behaved comparison is at least as common in the population.
\cite{mogstad2021causal} describe such a condition in the case of multiple instruments with a homogeneous treatment.
Additionally, in the case with multiple unobserved treatment varieties, it is possible that an individual who is a defier for one treatment is a complier for another (such individuals are the reason for complications beyond those for homogeneous treatments).
If such individuals have identical treatment effects for both versions of treatment, they are effectively the same as always-takers, or, phrased differently, they cancel out their own defier contribution to the estimand with their complier contribution to the estimand.%
\footnote{Selection on gains suggests some plausibility to this case. Individuals who are at the margin of two different treatment statuses may be marginal specifically because their treatment effects are similar.} 
The next section covers a setting when multiple instrumental variables are available to a researcher and only some satisfy the conditions described in this section.

\section{Factor-Augmented Treatment Effects}\label{section:FATE}
In this section, I describe identification and estimation of wage returns to college using a novel factor-augmented treatment effects model that leverages instruments that violate the uniform treatment responses condition above (and therefore also violate uniform unordered monotonicity) at the cost of a homogeneous treatment effects assumption on the unobserved component treatments.
The model imposes a factor structure on unobserved treatments and their effects, allowing for $K$ instruments to work through $L\leq K$ weighted averages of unobserved component treatments to affect $J$ lifecycle wage outcomes.%
\footnote{Though I consider a single outcome over many time periods in this application, the method described is not restricted to panel data.}
An intuitive interpretation of the model is that it produces data-driven lower-dimensional index variables for instruments, where each instrument's contributions to indexes are determined by its associated treatment effects across multiple outcomes.
When $L=1$, the model's treatment effect estimand for each outcome collapses to the 2SLS estimand.
When $L=K$, the model's treatment effect estimand for each outcome for unobserved factor $\ell$ collapses to the just-identified IV estimand for instrument $k=\ell$ for outcome $j$ while including all other IVs as controls.

I assume that data is generated from the following parametric model of wages ($Y_{ij}$ for time periods $j=1,2,...,J$) and unobserved component treatments ($D_{i\ell}$ for treatment components $\ell=1,2,...,L$),
\begin{equation}\label{FATE_DGP}
\begin{split}
Y_{ij} &= X_{i}\alpha_j+\sum_{\ell=1}^LD_{i\ell}\lambda_{\ell j}+\xi_{ij} \\
D_{i\ell} &= X_{i}\gamma_{x\ell}+\sum_{k=1}^Kz_{ik}\gamma_{k\ell}+u_{i\ell} 
\mbox{ }\mbox{ }\mbox{ }\mbox{ }\mbox{ }\mbox{ }\mbox{ }\mbox{ }\mbox{ }\mbox{ } \forall \mbox{ } \ell=1,2,...,L.
\end{split}
\end{equation}
$X_i$ contains an $N 
\times R$ vector of controls, and each $N \times 1$ instrument $z_{ik} \subset Z_i$ has marginal effect $\gamma_{k\ell}$ on unobserved component treatment $\ell$, while $\lambda_{\ell j}$ is the marginal effect of component treatment $D_{i\ell}$ as in the homogeneous treatment effects case in the preceding section. 
If we observed the component treatments, we could estimate model (\ref{FATE_DGP}) via 2SLS as long as there are at least as many instruments as component treatments $(K \geq L$).

We do not observe $D_{i\ell}$ for $\ell=1,2,...,L$, but we do observe $D_i$. It follows that we can estimate the following slightly unconventional IV model,
\begin{equation}\label{Reduced_form}
\begin{gathered}
Y_{ij} = X_{i}\beta_{j}+\sum_{k=1}^Kz_{ik}\gamma_k\pi_{kj}+\epsilon_{ij} \\
D_{i} =X_{i}\gamma_x + \sum_{k=1}^Kz_{ik}\gamma_k+ u_{i},
\end{gathered}
\end{equation}
where $\pi_{kj}$ is equivalent to the just-identified IV estimand of instrument $k$ for outcome $j$ when including all other instruments as controls, and $\gamma_k$ is the marginal effect of $z_{ik}$ on $D_i$.
Substituting the unobserved first stage into the unobserved second stage in model (\ref{FATE_DGP}) gives an equivalent observed second stage of,
\begin{equation}\label{Structural_Stage_2}
\begin{gathered}
Y_{ij}
= X_{i}(\alpha_j+\sum_{\ell=1}^L\gamma_{x\ell}\lambda_{\ell j})+\sum_{k=1}^Kz_{ik}\gamma_k\sum_{\ell=1}^L\frac{\gamma_{k\ell}}{\gamma_k}\lambda_{\ell j}+(\xi_{ij}+\sum_{\ell=1}^Lu_{i\ell}\lambda_{\ell j}) \\
= X_{i}\beta_j
+Z_i\gamma\theta\lambda_{j}
+\epsilon_{ij},
\end{gathered}
\end{equation}
where $Z_i = [z_{i1},z_{i2},...,z_{iK}]$, $\gamma = diag([\gamma_1,\gamma_2,...,\gamma_K])$, $\theta = [\theta_1, \theta_2,...,\theta_K]'$, with
$\theta_k = [\theta_{k1},\theta_{k2},...,\theta_{kL}]$ and $\theta_{k\ell} = \gamma_{k\ell}/\gamma_k$, and $\lambda_j = [\lambda_{1j},\lambda_{2j},...,\lambda_{Lj}]'$.
The reduced form coefficient on $Z_i$ is composed of $\gamma$ (which is identified from the first stage) and a weighted average of unobserved treatments' effects, $\theta \lambda_j$. 

Comparing equation (\ref{Structural_Stage_2}) with the second stage in model (\ref{Reduced_form}) reveals that the just-identified IV estimand for outcome $j$ using instrument $k$ has a factor structure, $\pi_{kj} = \theta_k \lambda_j$, where instrument $k$ places outcome-invariant weights, $\theta_k$, on instrument-invariant component treatment effects, $\lambda_j$.
The definition of $\theta_{k\ell}$ and the constraint $D_i=\sum_{\ell=1}^LD_{i\ell}$ implies that $\sum_{\ell=1}^L \theta_{k\ell}=1$, so the weights instrument $k$ places on unobserved factors must sum to one.
The identification of treatment effects is generally achieved via additional restrictions on $\theta$, some of which result in familiar estimands as we will see in the next subsection.

\subsection{Identification of Component Effects}\label{Identification}

To establish identification conditions for the factor-augmented treatment effects model, I make use of multiple outcomes and arrange each $\pi_{kj}$ into a $K \times J$ matrix $\pi$. 
The factor structure on instrument-specific treatment effects across many outcomes in (\ref{Structural_Stage_2}) implies that
\begin{equation}\label{Pi_structural}
\begin{split}
\pi = \theta \lambda,
\end{split}
\end{equation}
where the $K \times L$ matrix of weights instruments place on component treatments, $\theta$, is defined as above and the $L \times J$ matrix of component treatment effects across outcomes is given by $\lambda = [\lambda_1,\lambda_2,...,\lambda_J]$.
A barrier to identification, common to factor models, is that $\pi = \theta \lambda = \theta A A^{-1} \lambda$ for an arbitrary invertible $L \times L$ matrix $A$, so the factors and their weights are not separately identified even with many instruments and outcomes.

$A$ has $L^2$ free elements, so we require at least $L^2$ restrictions to identify the model.
Unlike many uses of factor models where the factors are effectively used as control variables, such as \cite{bai2009panel}, it is necessary for the current method that we preserve the economic interpretation and scale of the treatment effect parameters.
This adds some complications to identification, as many restrictions that will formally secure identification will nonetheless obscure the interpretation of model estimands.
To emphasize this, I denote by $(\theta^r,\lambda^r)$ the component weights and component treatment effects that satisfy identification restrictions.
Even in cases where all instruments are valid and $\pi_j$ can be interpreted as a weighted average of unobserved treatments' effects (as in Theorem \ref{theorem:UTR_effect}), it does not obviously follow that $\lambda^r_j$ is a weighted average of unobserved treatments' effects for all restrictions that formally secure identification of the model.

I propose the following restrictions for the model (though others are possible). 
First, I impose the normalization \hfill
\begin{equation}\label{Normalization}
    \theta^r = 
    \begin{bmatrix}
    I_L \\
    \theta^r_{L+1}\\
    \vdots \\
    \theta^r_K
    \end{bmatrix},
\end{equation}
which effectively defines the first $L$ component treatments in terms of the first $L$ instruments.
This restriction will ensure interpretability of $\lambda^r_\ell$ by tying it closely to the just-identified IV estimand for instrument $k=\ell$.
Because $D_i = \sum_{\ell=1}^LD_{i\ell}$ and the weights instruments place on component treatments are defined as $\theta_{k \ell} = \gamma_{k\ell}/\gamma_k$, it follows that $\sum_{\ell=1}^L\theta_{k \ell} = 1$. I therefore impose that \hfill
\begin{equation}\label{weights_sum_1}
    \sum_{\ell=1}^L\theta^r_{k \ell} = 1.
\end{equation}
Together, I describe $L^2+K-L$ restrictions, which is more than the formal requirement of $L^2$.
These restrictions clearly define the treatment effect estimands $\lambda^r$ while also imposing economically meaningful restrictions on the weights instruments place on unobserved treatments.

The relationships given in (\ref{Pi_structural}), (\ref{Normalization}), and (\ref{weights_sum_1}) provide $KJ$, $L^2$, and $K-L$ equations, respectively, where the first $L$ restrictions in (\ref{weights_sum_1}) are redundant given the restrictions in (\ref{Normalization}). 
Meanwhile, there are $KL$ parameters in $\theta^r$ and $JL$ in $\lambda^r$. 
It follows that identification is achieved if
\hfill
\begin{equation*}
    L^2+KJ+K-L \geq JL+KL,
\end{equation*}
where solving for $L$ when $K>$L yields
\hfill
\begin{equation}\label{Identification_condition}
    (K,J+1) \geq L,
\end{equation}
while the inequality is satisfied for any $J$ if $K=L$.
This expression defines the required number of outcomes and instruments for identification of a model with $L$ unobserved component treatments.

It remains to be shown that the identified parameter $\lambda^r$ defines a well-behaved treatment effect.
I show that $\lambda^r$ is a weighted average of unobserved component treatments even when some instruments violate uniform treatment responses.
\begin{thm}\label{theorem:UTR_FATE_effect}
    If the model in (\ref{FATE_DGP}) describes the data-generating process with $L$ unobserved component treatments where $L$ is known and there are $K\geq L$ instruments, Conditions 1-4 hold for all comparisons $(z_m,z_{-m})$ and $(z_m',z_{-m})$ in $\mathcal{Z}$ for all $m=1,2,...,K$, and Condition \ref{condition:uniform_treatment_responses} holds for $z_k$ with $k\leq L$, the following weighted average component treatment effect is identified: \hfill
    \begin{equation*}
    \lambda^r_{k j} =
    \sum_{\ell=1}^L 
    \theta_{k\ell}
    \lambda_{\ell j},
\end{equation*}
with $\theta_{k\ell} \geq 0$ for all $\ell$.
\end{thm}
\begin{proof}
Let Condition \ref{condition:uniform_treatment_responses} hold for instrument $k$ that satisfies $k\leq L$, with $P_\ell(z_k,z_{-k}) \geq P_\ell(z_k',z_{-k})$ and $z_k>z_k'$ for all $\ell$. 
If the model in (\ref{FATE_DGP}) describes the DGP, then it follows that $\gamma_k>0$ and $\gamma_{k \ell}>0$ for all $\ell$, such that $\theta_{k\ell} \geq 0$ for all $\ell$.
It also follows that the just-identified treatment effect estimands across all instruments for outcome $j$ are given by $\pi_j=\theta\lambda_j$.
The normalization given in (\ref{Normalization}) is equivalent to defining 
$\pi_j=\theta^r\lambda^r_j = \theta\theta_L^{-1}\theta_L\lambda_j$, where $\theta_L$ contains the first $L$ rows of $\theta$.
$\lambda^r_j=\theta_L\lambda_j$ implies that $\lambda^r_{k j} = \sum_{\ell=1}^L\theta_{k \ell}\lambda_{\ell j}$ for $k\leq L$, which implies that the normalized component treatment effect defined in terms of instrument $k$, $\lambda_{kj}^r$, is identified.
\end{proof}

The proof above requires that the number of unobserved treatment components be known to the researcher and be fewer than the number of instruments. 
It also requires homogeneous treatment effects for unobserved treatments' effects.
These are sufficient conditions for the result, and perhaps could be weakened.
Loosely speaking, if just-identified IV estimands for all instruments and all outcomes can be perfectly explained by constant weights on a lower-dimensional number of treatment effects, the result above will hold.
I leave consideration of possible weaker conditions under which the same or similar result holds for large numbers of unobserved treatments and/or heterogeneous effects of the unobserved component treatments to future work.

\subsection{Estimation}
This section describes estimation of the model.
Substituting the factor structure on treatment effects using $\pi_{kj} = \theta^r_{k\ell}\lambda^r_{\ell j}$ in model (\ref{Reduced_form}), expanding the model to contain equations for all outcomes, and rewriting it in matrix notation gives
\begin{equation}\label{Full_model}
\begin{split}
    Y_1 &= X\beta_1+Z\gamma\theta^r\lambda^r_1+\epsilon_1 \\
    Y_2 &= X\beta_2+Z\gamma\theta^r\lambda^r_2+\epsilon_2 \\
    &\vdots \\
    Y_J &= X\beta_J+Z\gamma\theta^r\lambda^r_J+\epsilon_J \\
    D &= X\gamma_x+Z\gamma+u.
 \end{split}   
\end{equation}
It is straightforward to estimate the above system via GMM with moments given by
\begin{equation}\label{GMM}
\begin{split}
    \mathbb{E}[Z^*(Y_1 &- X\beta_1-Z\gamma\theta^r\lambda^r_1)]=0 \\
    \mathbb{E}[Z^*(Y_2 &- X\beta_2-Z\gamma\theta^r\lambda^r_2)]=0 \\
    &\vdots \\
    \mathbb{E}[Z^*(Y_J &- X\beta_J-Z\gamma\theta^r\lambda^r_J)]=0 \\
    \mathbb{E}[Z^*(D &- X\gamma_x-Z\gamma)]=0,
 \end{split}   
\end{equation}
where $Z^* = [Z, X]'$.

It is helpful to consider two extreme cases to develop intuition for the estimator.
When $L=1$, the constraint in (\ref{weights_sum_1}) implies that $\theta$ is a $K \times 1$ vector of ones, such that the structural second stage in (\ref{Structural_Stage_2}) is equivalent to the second stage in (\ref{Reduced_form}) under the single treatment effect constraint $\pi_{kj}=\pi_j$, where $\pi_j = \lambda_j$ is the scalar IV-GMM estimand.
When $L=K$, the constraint in (\ref{weights_sum_1}) implies that $\theta$ is a $K \times K$ identity matrix, such that the structural second stage in (\ref{Structural_Stage_2}) is identical to the second stage in (\ref{Reduced_form}) with $\lambda_j$ = $\pi_j = [\pi_{1j},\pi_{2j},...,\pi_{Kj}]'$.
The factor-augmented treatment effects model thus nests IV-GMM  and just-identified IV as special cases.
For $L<K$, the model allows many instruments to identify a lower-dimensional number of effects for each outcome, with the goal of achieving efficiency gains similar to those that may be achieved by 2SLS or IV-GMM without requiring every instrument to satisfy uniform treatment responses.

\subsection{Testing the Number of Component Treatments}

The previous section describes estimation of the FATE model for $L$ unobserved component treatments, where $L$ is known.
This section describes a statistical test for the null hypothesis that the model is correctly specified, which means instruments satisfy independence, exclusion, and uniform treatment responses, with each of $L$ unobserved treatments having homogeneous effects. 
If this test is rejected and the researcher is sure that the other conditions are satisfied, it is possible that the instruments work through more than $L$ unobserved component treatments to drive outcomes.

If the above assumptions are satisfied and as $N \rightarrow \infty$, the moment conditions in (\ref{GMM}) will hold exactly.
In small samples, these moment conditions might not exactly hold, but we can test the hypothesis that they hold in the population by replacing them with within-sample estimates:
\begin{equation}\label{GMM_sample}
\begin{split}
    \sum_{i=1}^NZ_i^*(Y_{i1} &- X_i\hat\beta_1-Z_i\hat\gamma\hat\theta^r\hat\lambda^r_1)=0 \\
    \sum_{i=1}^NZ_i^*(Y_{i2} &- X_i\hat\beta_2-Z_i\hat\gamma\hat\theta^r\hat\lambda^r_2)=0 \\
    &\vdots \\
    \sum_{i=1}^NZ_i^*(Y_{iJ} &- X_i\hat\beta_J-Z_i\hat\gamma\hat\theta^r\hat\lambda^r_J)=0 \\
    \sum_{i=1}^NZ_i^*(D_i &- X_i\hat\gamma_x+Z_i\hat\gamma)=0,
 \end{split}   
\end{equation}
The implied test is the \cite{hansen1982large} J-test, which tests the null that the sample estimates of system (\ref{GMM}) are satisfied with estimated values.
When the assumed value $L$ is $L=1$, this is similar to the J-test of overidentifying restrictions for IV-GMM.
When $L=K$, the test is trivially satisfied because the model is estimated to set the sample moments to zero.
When $L$ is between $1$ and $K$, a large $J$-statistic indicates that an identification assumption is likely to be violated, which includes the possibility that the true number of unobserved treatments is larger than the assumed value of $L$.

\section{Data}
I estimate returns to college with the factor-augmented treatment effects model described above using data from the National Longitudinal Survey of Youth 1979 (NLSY79).
The NLSY79 follows a group of 12,686 individuals who were between the ages of 14 and 22 in 1979. Respondents were surveyed annually for the first 15 years and biennially afterwards. Survey responses include information on years of education, earnings, geography of residence, and a wealth of auxiliary information. In addition to meeting the requirements of my model of containing multiple instruments and outcomes, this dataset is attractive because it has been used extensively in the literature on returns to education, which helps to emphasize my methodological contributions by holding fixed some other potential explanations for similarities and differences in results. In particular, I use the same sample of white males that was used by \cite{carneiro2011estimating}, matched with longitudinal wage data constructed according to the procedure described in \cite{ashworth2021changes}.

The composite treatment of interest is college attendance as of age 25. Estimating my model requires multiple outcomes and instruments, as shown in expression (\ref{Identification_condition}). The longitudinal nature of the NLSY allows me to set $j=t$ for $t=26,\dots,50$ and rely on wage outcomes over the lifecycle to identify the model.%
\footnote{Researchers considering using this method in other applications should rarely be deterred by a lack of outcomes. Even if economically meaningful outcomes are sparse in a given dataset, it is possible to estimate the model using outcomes that lack obvious economic significance but nonetheless contribute to identification.}

I  use several instruments that are established in the literature on returns to education. They are average wages in the county of residence at age 17, an indicator for the presence of a college in county of residence at age 14, and average tuition of public colleges in the county of residence at age 17.
\footnote{
The unemployment rate in the state of residence at age 17 is another commonly used instrument. I have opted to exclude it because it is extremely weak compared to the other instruments.
}
That each of these variables contributes to the college attendance decision is intuitive as well as testable. I discuss the validity of each instrument separately in terms of independence, exclusion, and uniform treatment responses.

Local wages have been used as explanatory variables for educational decisions by \cite{cameron1998life}, \cite{cameron2001dynamics}, \cite{cameron2004estimation}, and \cite{carneiro2011estimating}. 
Following \cite{carneiro2011estimating}, I use the average values of this variable at age 17 at the county level as an instrument for college attendance. 
Because local labor market conditions may contribute to outcomes through channels other than education, I control for permanent wages in the county of residence at age 17, permanent unemployment in the state of residence at age 17, wages in the county of residence in 1991, and unemployment in the state of residence in 1991 (to control for persistent early life-cycle labor market shocks).
It follows that I am effectively using temporal deviations from permanent local wages at age 17 as an instrument. 
The primary channel through which I expect this instrument to affect college attendance is by making the outside option of directly entering the workforce more attractive, without altering individuals' preferences over types of college, suggesting that it is particularly likely to satisfy uniform treatment responses (and uniform unordered monotonicity).
It is also possible that this instrument affects college decisions by relaxing credit constraints, allowing individuals who would not have attended college to do so, and allowing individuals who would have attended college to attend more expensive colleges, though this explanation is at odds with the empirical finding of a negative first stage coefficient for this instrument.

The presence of a college in the county of residence at age 14 has been used as an instrument for education by \cite{card1993using, kane1995labor, kling2001interpreting, currie2003mother, cameron2004estimation, carneiro2011estimating}, and \cite{mogstad2021causal}. 
I expect the presence of colleges in the county of residence to primarily influence education by encouraging students to attend a nearby college through a cost reduction (both pecuniary and psychic) mechanism. 
If this sort of cost reduction is more important for students inclined toward particular types of college education, such as majors \citep{altonji2012heterogeneity} or quality \citep{black2006estimating}, then this IV will identify the effects of such types of education, and may violate uniform treatment responses if individuals prefer nearby colleges of an otherwise non-preferred types to distant colleges of more-preferred types.
For instance, if college returns are only a function of quality and some individuals induced into college by this instrument have a preference ordering of nearby high-quality college $\succ$ nearby low-quality college $\succ$ distant high-quality college, we might expect this instrument to partially represent the difference between returns to low-quality college and high-quality college (driven by such individuals whose nearby college is low-quality).
Independence and exclusion for this instrument are threatened by correlation between college presence and students' ability, as noted by \cite{carneiro2002evidence} and \cite{cameron2004estimation}. 
To address this, I control for a measure of ability (the Armed Forces Qualification Test, AFQT), following \cite{carneiro2011estimating}.

Local tuition costs have been used as an instrument for education by \cite{kane1995labor}, \cite{cameron1998life}, \cite{cameron2001dynamics}, and \cite{carneiro2011estimating}. 
This is similar to local labor market conditions in that it may be correlated with unobserved ability, such as if expensive colleges produce different externalities for the local community than cheap colleges. 
I control for mother's years of education, permanent local labor market conditions, and AFQT to address this concern. 
I view this instrument as a pecuniary cost-shifter, which produces particular concerns regarding uniform treatment responses.
Cost-shifter instruments produce income effects, which I expect to induce exposed individuals to obtain leisure, potentially by putting less effort into their studies.
If low-effort college has lower returns than high-effort college and this instrument shifts sufficiently many individuals from high-effort college to low-effort college, it will place negative weights on high-return college effects, yielding a downward-biased estimate.
For this reason, I choose this instrument to not identify an unobserved treatment variety, explicitly allowing it to contribute to estimation while taking on negative weights.

I  use several variables as controls, which is common in the literature. They are number of siblings, 
permanent local unemployment in the state of residence at age 17 (average from years 1979-2000), permanent local log wages in the state of residence at age 17 (average from years 1979-2000), a binary indicator for urban residence at age 14, 
average wages in the county of residence in 1991, average unemployment in the county of residence in 1991, 
year of birth indicators, mother's years of education, and the Armed Forces Qualifying Test score. Statistics on the variables I use are shown in Table \ref{tab:descriptives}.

\begin{table}[htbp]\centering
\def\sym#1{\ifmmode^{#1}\else\(^{#1}\)\fi}
\caption{Summary Statistics}
\label{tab:descriptives}
\begin{tabular}{l*{3}{c}}
\midrule
& \multicolumn{1}{c}{Mean} & \multicolumn{1}{c}{Standard deviation} \\
& \multicolumn{1}{c}{(1)} & \multicolumn{1}{c}{(2)} \\
\midrule 
Mean log wages, ages 26-30 & 2.084 & 0.463 \\
Mean log wages, ages 31-35 & 2.172 & 0.533 \\
Mean log wages, ages 36-40 & 2.201 & 0.575 \\
Mean log wages, ages 41-45 & 2.198 & 0.593 \\
Mean log wages, ages 46-50 & 2.156 & 0.618 \\
Attended college & 0.495 & 0.500 \\
Number of siblings & 2.927 & 1.909 \\
Permanent local unemployment at age 17 & 6.251 & 0.986 \\
Permanent local log wages at age 17 & 10.283 & 0.188 \\
Urban residence at age 14 & 0.744 & 0.436 \\
Local unemployment rate in 1991 & 6.810 & 1.267 \\
Local log wages in 1991 & 10.293 & 0.165 \\
Year of birth & 1959.759 & 2.340 \\
Mother's years of schooling & 12.102 & 2.335 \\
Corrected AFQT score & 0.449 & 0.952 \\
Local log wages at age 17 & 10.276 & 0.164 \\
Nearby four-year college at age 14 & 0.525 & 0.500 \\
Local tuition at age 17 & 21.568 & 7.981 \\
\midrule  
\multicolumn{1}{c}{Sample size}&\multicolumn{2}{c}{1747}  \\
\bottomrule  

\end{tabular}
\begin{minipage}{1\linewidth}
\smallskip
\footnotesize
\emph{Notes:} Means and standard deviations for white males in the National Longitudinal Survey of Youth 1979 sample. AFQT corrected for years of schooling at the timing of test-taking. Permanent log wages and unemployment are calculated as average values from 1973 to 2000 for the county of residence at age 17. 
\end{minipage}
\end{table}

\section{Results}

\begin{figure}[hbtp!]
\centering
\includegraphics[width=.8\linewidth]{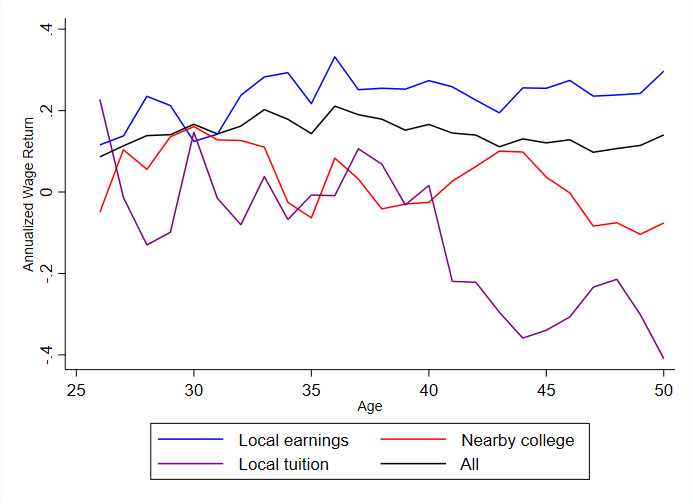}
\begin{minipage}{1\linewidth}
\caption{IV-GMM Estimates of Lifecycle Wage Returns to College}
\label{fig:TSLS}
\smallskip
\footnotesize
\emph{Notes:} This graph depicts IV-GMM point estimates of the effects of college attendance on lifecycle log wages for each excluded instrument listed, as well as the case where all instruments are excluded. See Table \ref{tab:TSLS} for point estimates and standard errors.
\end{minipage}
\end{figure}

 I  begin by showing IV treatment effects estimated via GMM for just-identified IV excluding each instrument individually and including the others as controls, as well as IV-GMM excluding all IVs from outcome equations estimation. 
 These results are in Table \ref{tab:TSLS}, and are shown graphically in Figure \ref{fig:TSLS}. 
 I divide all coefficients by four to express returns in annual terms.
 This table reports F-statistics for each model's first stage, as well as p-values for the Hansen J test of overidentifying restrictions for the model excluding all IVs from outcome equations.

The first stage F-stats for local wages, nearby college, and local tuition when used alone are 18.7, 5.5, and 2.5, respectively, with the F-stat for all of them is 11.1.
Two of these fall below the commonly-invoked heuristic threshold of 10, while also falling short of the corresponding statistics calculated in a similar exercise by \cite{mogstad2021causal}, likely due to correlation between IVs and my inclusion of a third that they do not use (local tuition).
The particularly weak instruments, nearby college and local tuition, have intuitive (positive and negative, respectively) signs on college attendance.
\cite{angrist2021one} show that sign-screening reduces bias from weak IVs, so I assume that the instruments are immune to weak IV problems.%
\footnote{Because I would have used IVs common to the literature (and this dataset) regardless of their first stages, it is inaccurate to say that I sign-screened these. It is likely, however, that prior researchers sign-screened these IVs when introducing them, so the sign-screening argument holds.} 

The local earnings instrument consistently estimates relatively high returns to college, which grow larger later in the lifecycle.
I argued above that this instrument is particularly likely to satisfy net uniform unordered monotonicity, suggesting that these estimates likely place positive weight on returns to different types of college for different types of students.
It follows that college may be a better investment than is sometimes suggested in the literature, with many estimates suggesting returns in the proximity of 10\% \citep{card2001estimating}.
It is possible that other instruments are valid as well and this instrument places particularly high weight on high-return versions of college.
This could occur if particularly money-motivated individuals, who are inclined toward high-paying versions of college, are particularly sensitive to local variation in earnings when they make their education decisions.
In either case, the returns implied by this instrument suggest that extremely high return versions of college exist for some individuals, which could suggest scope for substantial social gains by finding ways to induce more individuals to attend such versions of college.

The nearby college and local tuition instruments both run some risk of inducing students to shift out of high-effort college and toward low-effort college, as they both can be seen as cost-shifters which likely have income effects.
It follows that the estimates for these instruments should be lower than true returns if low-effort college has lower returns than high-effort college, which is consistent with these point estimates being consistently lower throughout the lifecycle than those identified using the local earnings IV.
That said, the standard errors on estimates from both of these IVs are quite large, which makes it difficult to draw sharp conclusions.

Excluding all IVs from outcome equations, as in model (4), yields substantial efficiency gains as evidenced by the substantially smaller standard errors.
However, it is possible that these estimates inherit some of the bias from net uniform unordered monotonicity violations of the nearby college and local tuition instruments.
Noteably, the point estimates are almost always smaller when using all IVs to estimate treatment effects relative to only using the local earnings IV, which is consistent with the possibility of these estimates shifting some individuals out of high-return college and into low-return college.

\begin{table}[htbp]\centering
\def\sym#1{\ifmmode^{#1}\else\(^{#1}\)\fi}
\caption{IV-GMM Estimates of Lifecycle Returns to College, by IV}
\label{tab:TSLS}
\begin{tabular}{@{\extracolsep{4pt}}l*{8}{c}}
\hline
& \multicolumn{8}{c}{2nd Stage Excluding:} \\
\cline{2-9} 
& \multicolumn{2}{c}{Local earnings}& \multicolumn{2}{c}{Nearby college}& \multicolumn{2}{c}{Local tuition}& \multicolumn{2}{c}{All}\\ 
& \multicolumn{2}{c}{(1)}& \multicolumn{2}{c}{(2)}& \multicolumn{2}{c}{(3)}& \multicolumn{2}{c}{(4)}\\ 
 \hline 
Log wage, age 26  & 0.116 & (0.086) & -0.050 & (0.145) & 0.227 & (0.226) & 0.087 & (0.052) \\
Log wage, age 27  & 0.138 & (0.074) & 0.104 & (0.142) & -0.013 & (0.194) & 0.113 & (0.053) \\
Log wage, age 28  & 0.235 & (0.087) & 0.056 & (0.131) & -0.130 & (0.216) & 0.139 & (0.054) \\
Log wage, age 29  & 0.213 & (0.083) & 0.136 & (0.139) & -0.099 & (0.199) & 0.141 & (0.055) \\
Log wage, age 30  & 0.125 & (0.092) & 0.161 & (0.136) & 0.147 & (0.210) & 0.167 & (0.056) \\
Log wage, age 31  & 0.142 & (0.091) & 0.128 & (0.133) & -0.015 & (0.200) & 0.143 & (0.054) \\
Log wage, age 32  & 0.238 & (0.082) & 0.127 & (0.134) & -0.080 & (0.211) & 0.162 & (0.054) \\
Log wage, age 33  & 0.283 & (0.094) & 0.110 & (0.142) & 0.038 & (0.200) & 0.203 & (0.060) \\
Log wage, age 34  & 0.293 & (0.102) & -0.025 & (0.144) & -0.067 & (0.212) & 0.179 & (0.066) \\
Log wage, age 35  & 0.217 & (0.118) & -0.063 & (0.151) & -0.007 & (0.203) & 0.143 & (0.069) \\
Log wage, age 36  & 0.332 & (0.109) & 0.083 & (0.141) & -0.009 & (0.199) & 0.211 & (0.065) \\
Log wage, age 37  & 0.251 & (0.099) & 0.032 & (0.139) & 0.106 & (0.190) & 0.190 & (0.062) \\
Log wage, age 38  & 0.255 & (0.099) & -0.041 & (0.152) & 0.068 & (0.188) & 0.179 & (0.064) \\
Log wage, age 39  & 0.253 & (0.091) & -0.029 & (0.149) & -0.031 & (0.203) & 0.152 & (0.060) \\
Log wage, age 40  & 0.274 & (0.094) & -0.025 & (0.148) & 0.016 & (0.188) & 0.166 & (0.060) \\
Log wage, age 41  & 0.259 & (0.095) & 0.027 & (0.149) & -0.219 & (0.272) & 0.145 & (0.059) \\
Log wage, age 42  & 0.226 & (0.093) & 0.063 & (0.151) & -0.222 & (0.259) & 0.140 & (0.060) \\
Log wage, age 43  & 0.195 & (0.087) & 0.100 & (0.151) & -0.295 & (0.299) & 0.111 & (0.059) \\
Log wage, age 44  & 0.256 & (0.089) & 0.098 & (0.145) & -0.359 & (0.334) & 0.131 & (0.058) \\
Log wage, age 45  & 0.255 & (0.089) & 0.037 & (0.147) & -0.339 & (0.332) & 0.121 & (0.057) \\
Log wage, age 46  & 0.274 & (0.106) & -0.002 & (0.148) & -0.307 & (0.319) & 0.128 & (0.059) \\
Log wage, age 47  & 0.236 & (0.102) & -0.084 & (0.158) & -0.233 & (0.282) & 0.098 & (0.059) \\
Log wage, age 48  & 0.238 & (0.108) & -0.075 & (0.157) & -0.214 & (0.269) & 0.107 & (0.060) \\
Log wage, age 49  & 0.242 & (0.094) & -0.104 & (0.167) & -0.301 & (0.331) & 0.115 & (0.064) \\
Log wage, age 50  & 0.297 & (0.108) & -0.076 & (0.168) & -0.409 & (0.383) & 0.140 & (0.069) \\
 \hline
 First Stage F-stat & 18.709 &  & 5.220 &  & 2.472 & & 11.126 \\
Over-id p-value & & & & & & & 0.339 \\
 Sample size & 1747 &  & 1747 &  & 1747 & & 1747 \\

\hline
\hline
\end{tabular}
\begin{minipage}{1\linewidth}
\smallskip
\footnotesize
\emph{Notes:} Two-step IV generalized method of moments  estimates of the effects of college attendance on log wages for white men in the NLSY79 over the lifecycle. Age-specific estimates for models 1-3 obtained excluding a single instrument and including all others as controls along with the other control variables listed in Table \ref{tab:descriptives}. Model 4 excludes all instruments from outcome equations. College returns are annualized by dividing the college attendance coefficient by four. Robust standard errors in parentheses. 
\end{minipage}
\end{table}

\begin{table}[htbp]\centering
\def\sym#1{\ifmmode^{#1}\else\(^{#1}\)\fi}
\caption{GMM Estimates of FATE Model of Lifecycle Returns to College}
\label{tab:UTH}
\begin{tabular}{l*{7}{c}}
\hline
& \multicolumn{2}{c}{Local earnings component}& \multicolumn{2}{c}{Nearby college component}\\ 
& \multicolumn{2}{c}{(1)}& \multicolumn{2}{c}{(2)}\\ 
\hline 
Component effects ($\lambda^r$) \\
Log wage, age 26  & 0.076 & (0.074) & 0.089 & (0.055) \\ 
Log wage, age 27  & 0.143 & (0.073) & 0.105 & (0.058) \\ 
Log wage, age 28  & 0.234 & (0.085) & 0.136 & (0.076) \\ 
Log wage, age 29  & 0.214 & (0.082) & 0.137 & (0.068) \\ 
Log wage, age 30  & 0.130 & (0.086) & 0.134 & (0.059) \\ 
Log wage, age 31  & 0.136 & (0.085) & 0.099 & (0.060) \\ 
Log wage, age 32  & 0.227 & (0.078) & 0.150 & (0.067) \\ 
Log wage, age 33  & 0.259 & (0.088) & 0.197 & (0.070) \\ 
Log wage, age 34  & 0.278 & (0.096) & 0.179 & (0.079) \\ 
Log wage, age 35  & 0.173 & (0.097) & 0.121 & (0.079) \\ 
Log wage, age 36  & 0.301 & (0.097) & 0.209 & (0.080) \\ 
Log wage, age 37  & 0.227 & (0.089) & 0.178 & (0.069) \\ 
Log wage, age 38  & 0.234 & (0.091) & 0.171 & (0.073) \\ 
Log wage, age 39  & 0.231 & (0.085) & 0.137 & (0.077) \\ 
Log wage, age 40  & 0.244 & (0.087) & 0.162 & (0.074) \\ 
Log wage, age 41  & 0.244 & (0.088) & 0.122 & (0.082) \\ 
Log wage, age 42  & 0.221 & (0.086) & 0.114 & (0.078) \\ 
Log wage, age 43  & 0.199 & (0.082) & 0.081 & (0.081) \\ 
Log wage, age 44  & 0.245 & (0.085) & 0.097 & (0.092) \\ 
Log wage, age 45  & 0.238 & (0.083) & 0.091 & (0.091) \\ 
Log wage, age 46  & 0.261 & (0.100) & 0.111 & (0.095) \\ 
Log wage, age 47  & 0.218 & (0.096) & 0.085 & (0.090) \\ 
Log wage, age 48  & 0.210 & (0.102) & 0.083 & (0.089) \\ 
Log wage, age 49  & 0.204 & (0.085) & 0.055 & (0.100) \\ 
Log wage, age 50  & 0.259 & (0.097) & 0.075 & (0.114) \\ 
 \hline
Component weights ($\theta^r$) \\
Local earnings &     1 & (.) &     0 & (.) \\ 
Nearby college &     0 & (.) &     1 & (.) \\ 
Local tuition & -2.498 & (2.333) & 3.498 & (2.333) \\ 
 \hline 
Value of criterion function (Q) & 0.009 \\
Over-id p-value & 0.904 \\
 Sample size & 1747 \\
 \hline \hline 

\end{tabular}
\begin{minipage}{1\linewidth}
\smallskip
\footnotesize
\emph{Notes:} Two-step generalized method of moments estimates of the factor-augmented treatment effect model of effects of college attendance on log wages for white men in the NLSY79. Component-defining instruments have weights normalized to 1 on respective components. College returns are annualized by dividing the college component effects by four. Robust standard errors in parentheses. 
\end{minipage}
\end{table}

\begin{figure}[hbtp!]
\centering
\includegraphics[width=.8\linewidth]{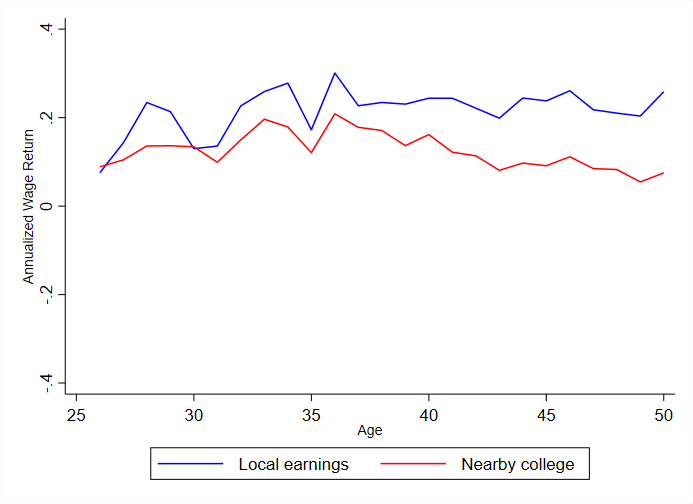}
\begin{minipage}{1\linewidth}
\caption{GMM Estimates of Lifecycle Wage Returns to Unobserved College Components}
\label{fig:Lambda_hat}
\smallskip
\footnotesize
\emph{Notes:} This graph depicts GMM point estimates of treatment effects from the factor-augmented treatment effects model on lifecycle log wages with components defined in terms of listed instruments. See Table \ref{tab:UTH} for point estimates and standard errors.
\end{minipage}
\end{figure}

I also estimate the factor-augmented treatment effects model in (\ref{Full_model}) with results shown in Table \ref{tab:UTH}, defining the first unobserved treatment in terms of the local earnings IV.
From theorem \ref{theorem:UTR_effect}, this will provide consistent estimates of a positively-weighted average of unobserved treatments' effects if the local earnings IV satisfies independence, exclusion, relevance, uniform treatment responses, and I properly specify the number of unobserved treatment varieties.
I define the second component in terms of the nearby college instrument.
If nearby college satisfies the required conditions, these estimates are consistent for a weighted average of effects of unobserved component treatments, whereas the estimates in column (2) are nuisance parameters if local earnings is a valid IV and nearby college is not.
Because of the concerns with nearby college listed above, I opt to view the associated estimates as nuisance parameters, but the reader is free to make their own decision.

The F-stat of 18.7 in Table \ref{tab:TSLS} suggests relevance is satisfied for local earnings, while the p-value on the test of overidentifying restrictions of 0.339 in Table \ref{tab:UTH} means that we fail to reject the null that $L=2$ component treatments are sufficient to explain the data.
The FATE model allows extra instruments to contribute to estimation with negative weights, which we see in the case of local tuition placing weights of -2.5 and 3.5 on the local earnings and nearby college treatment components, respectively.
As in the IV-GMM results, the local earnings component treatment has substantially higher point estimates through the lifecycle than the nearby college component.
Interestingly, the local earnings component point estimates visually look (both in the table and in Figure \ref{fig:Lambda_hat}) like those of just-identified IV using the local earnings instrument, while the nearby college component effects look like those of IV-GMM using all instruments for estimation of treatment effects.

Perhaps for similar reasons, the standard errors for the (potentially uninterpretable) nearby college component treatment effect estimates are dramatically smaller than those for the nearby college instrument in the standard IV-GMM results in Table \ref{tab:TSLS}, while the standard errors for the local earnings component are essentially unchanged (with trivial reductions).
Possible intuition for this is that the FATE model was able to rationalize the just-identified IV estimates of the local tuition instrument by altering the nearby college component treatment effects, without needing to alter the local earnings component treatment effects very much relative to the just-identified IV estimates.
It follows that the local tuition IV helps to estimate the nearby college component treatment effect parameters without lending much to the local earnings component effect parameter estimation.
This is economically reasonable insofar as both nearby college and local tuition may work through a cost-shifter channel, inducing similar types of individuals into similar types of college (with local tuition perhaps doing so more extremely, as evidenced by it placing a weight of 3.5 on the nearby college treatment component).

\section{Conclusion}

A large literature describes implications of unmodeled treatment effect heterogeneity for estimation of treatment effects.
This paper addresses the related problem of unmodeled treatment heterogeneity for treatment effect estimation. 
It shows the validity of standard methods when instruments shift types of individuals into (and not between) diverse varieties of treatment on net, as well as providing an alternative factor-augmented treatment effects model that allows instruments that shift individuals between types of treatment to contribute to treatment effect estimation.
I estimate the FATE model in the context of returns to college using the NLSY79, finding particularly high returns to college that range between 7.6\% and 30.1\% over the lifecycle for the component treatment associated with the instrumental variable least-likely to shift individuals between treatments (local earnings at age 17).

For empirical applications of instrumental variables methods that make use of single instruments, the most relevant contribution of this paper is the assumption of net uniform unordered monotonicity, which allows for unmodeled versions of treatment that affect outcomes differently to reside in the residuals of outcome equations (formally violating the exclusion restriction for modeled composite treatments).
As in the case of discussions of the monotonicity condition of \cite{imbens1994identification}, I recommend that empirical researchers justify the net uniform unordered monotonicity assumption when possible, or discuss its implications for their estimates when it may not be valid.
For instance, an instrument that violates net uniform monotonicity will be only minimally biased if there is minimal heterogeneity in effects between different versions of treatment.
Furthermore, treatment effects estimated using a policy-relevant instrument may be informative for changes in that particular policy even if it violates net uniform unordered monotonicity.
For instance, the returns to college estimated using local tuition variation may be relevant for describing the likely effects on earnings of increases in subsidization of college, if such counterfactual subsidies move individuals between versions of treatments the same way the instrument does.

For empirical applications involving multiple instruments, the factor-augmented treatment effects model developed here provides an alternative to 2SLS or other standard methods for pooling variation from multiple instruments into a single treatment effect estimate.
The FATE method is likely to be particularly useful when a particularly credible instrument, such as randomized assignment to treatment, is relatively weak, while some less credible instruments that may violate net uniform unordered monotonicity are relatively strong.

An attractive option for this is to construct a binary indicator set to one if an instrument takes the value with the lowest propensity of treatment, and use this constructed instrument to define a component treatment in the FATE model while allowing the continuous (or other functional form) versions of the instrument to place negative weights on unobserved versions of treatment.
In the extreme case where the treatment propensity is zero at a particular value of the instrument, it is impossible for there to be any violations of net uniform unordered monotonicity, and such violations are likely to still be rare when the treatment propensity is close to zero.
Such an exercise can be implemented in applications with a single strong instrument that might not satisfy net uniform unordered monotonicity when used as a single continuous variable, but which retains sufficient strength for treatment effect estimation when split into multiple instruments.

The sufficient conditions given above for the factor-augmented treatment effects model to identify convex averages of unobserved treatments' effects do not allow for individual-level treatment effect heterogeneity.
Some heterogeneity in treatment effects between individuals can be attributed to heterogeneity in chosen treatments, but it is plausible that a small number of unobserved treatments (two in the application) is insufficient to fully characterize the relevant heterogeneity.
Future theoretical work on the factor-augmented treatment effects estimator to determine its validity in the presence of treatment effect heterogeneity and/or more unobserved component treatments than instruments may be helpful in increasing its usefulness for applications.

In terms of the application to returns to college, I find evidence that effects of college on wages are potentially higher than are commonly found in the literature.
Estimating returns as driven by local wages at age 17, which is unlikely to alter students' preferences over different types of college, gives treatment effects from 12.5\% to 33.2\% (IV-GMM) or 7.6\% to 30.1\% (FATE), which are generally much higher than the returns around 10\% that are commonly found in the literature.
Insofar as variation in returns between instruments is due to instruments inducing individuals into different types of treatments, these results suggest substantial potential for gains (in terms of wage increases) from policies that move students into higher-return types of college.
Insofar as students' current choices are driven in part by information frictions regarding pecuniary returns to college or students internalize nonpecuniary benefits from types of college more than pecuniary benefits (for instance due to taxes), aggregate welfare gains are likely as well.

The distinction between heterogeneous treatment effects and heterogeneous treatments is central to these policy implications.
If some students have high returns and others have low returns, income-maximizing (or similar) policies are those that induce students to attend college if and only if they have positive returns.
If, however, different versions of college have different effects (in line with a very large literature on college majors and quality), then optimal policy will induce particular students into well-matched versions of college.
The finding, such as in \cite{cunha2005separating} or \cite{carneiro2011estimating}, that a large minority of individuals have negative returns to college is potentially due to students failing to choose versions of college that best-improve their human capital.
Further work to distinguish between heterogeneous effects of treatments and effects of heterogeneous treatments is essential to inform policies that seek to improve educational allocations.

\clearpage

\bibliographystyle{econometrica.bst}
\renewcommand{\bibname}{BIBLIOGRAPHY}
\bibliography{References}


\appendixpageoff
\begin{appendices}
\renewcommand{\thesection}{Appendix \Alph{section}:}
\renewcommand{\thesubsection}{\Alph{subsection}.\arabic{subsection}}

\section{Factor Augmented Treatment Effects Example}\label{Appendix:Examples}

In this section, I walk through an intuitive three-step estimation procedure for an extreme case of the model where there are three instruments ($K=3)$, two unobserved treatment varieties ($L=2$), and $J$ outcomes, where $J\geq 2$. 
The first step is to estimate just-identified IV treatment effects for each instrument for all outcomes using standard methods, yielding $\hat \pi$.
The second step infers $(\hat \lambda_j, \hat\theta)$ from $\hat\pi_j$ for a single outcome.
The third step takes $(\hat\theta)$ as given (similarly to 2SLS implicitly imposing ex ante that all instruments weight a single common version of treatment), yielding $\hat \lambda_{-j}$ for all outcomes other than $j$.

Considering only outcome $j$, the restrictions in (\ref{Pi_structural}), (\ref{Normalization}), and (\ref{weights_sum_1}) set 
\begin{equation*}
\begin{split}
    \hat\pi_{1,j} &= \hat\lambda_{1,j} \\
    \hat\pi_{2,j} &= \hat\lambda_{2,j} \\
    \hat\pi_{3,j} &= \hat\theta_{3,1}\hat\lambda_{1,j}+ (1-\hat\theta_{3,1})\hat\lambda_{2,j},
\end{split}
\end{equation*}
which is a system of three equations with three unknowns.
It follows that the unobserved treatments' effects and their outcome-invariant weights are exactly identified.

The weight parameters in $\hat\theta$ estiamted using outcome $j$ can be taken as given for all other outcomes.
It follows that estimates of $(\lambda_{1,j'},\lambda_{2,j'})$ can then be obtained as the two parameters that best fit the system of three equations,
\begin{equation*}
\begin{split}
    \hat\pi_{1,j'} &= \hat\lambda_{1,j'} \\
    \hat\pi_{2,j'} &= \hat\lambda_{2,j'} \\
    \hat\pi_{3,j'} &= \hat\theta_{3,1}\hat\lambda_{1,j'}+ (1-\hat\theta_{3,1})\hat\lambda_{2,j'},
\end{split}
\end{equation*}
for any outcome $j' \neq j$.
When $N\rightarrow\infty$ and there is no model misspecification, this seemingly overidentified system of equations will hold exactly.

In the example given, only a single outcome is used to estimate weights, and they are taken as given for all other outcomes.
In the GMM estimator described in Section \ref{section:FATE}, all outcomes contribute to estimation of weights.
Intuitively, in the case where $J \rightarrow \infty$, each outcome's contribution to the estimation of weights goes to zero, and the simultaneously-estimated model takes on more of the intuition described in the final step here.
Similarly, if a researcher has access to auxiliary outcomes that are not of substantive interest to the research question, it is possible to construct an estimator that estimates instrument weights on unobserved treatments using the auxiliary treatments only, while taking the weights as given for the outcomes of interest.%
\footnote{The use of auxiliary outcomes is similar to the use of test equations by \cite{cunha2005separating} and \cite{sarzosa2015bullying}, among others, when estimating models that control for unobserved heterogeneity in ability.}
Such an estimator could be implemented in steps as described here, or could be implemented via imposing a GMM weighting matrix with arbitrarily high weights on auxiliary outcomes, forcing the estimator to choose $\theta$ that prioritizes the fit for those particular moments, leaving other moments to take the implied $\hat \theta$ as given.

\end{appendices}

\end{document}